\documentclass[aps,prl,twocolumn,showpacs,
superscriptaddress, groupedaddress,nofootinbib]{revtex4}  % for review and submission
\usepackage{graphicx}  % needed for figures
\usepackage{dcolumn}   % needed for some tables
\usepackage{bm,relsize}        % for math
\usepackage{amssymb, amsmath}
\usepackage{slashed}   % for math

\usepackage[usenames,dvipsnames,svgnames,table]{xcolor}

\def\beq{\begin{equation}}
\def\eeq{\end{equation}}
\def\MDM	{M_{\mathsmaller{\rm DM}}}
\def\DM		{\mathsmaller{\rm DM}}
\def\se		{\sigma_e}

\begin{document}
% The following information is for internal review, please remove them for submission
\widetext
\leftline{DESY 18-194}
\leftline{KEK-TH 2085}

%\centerline{NOT FOR PUBLIC DISTRIBUTION}

\title{\Large Light Dark Matter at Neutrino Experiments}
\author{Yohei Ema}
\affiliation{DESY, Notkestra{\ss}e 85, D-22607 Hamburg, Germany}
\affiliation{KEK Theory Center, Tsukuba 305-0801, Japan}
\author{Filippo Sala}
\affiliation{DESY, Notkestra{\ss}e 85, D-22607 Hamburg, Germany}
\author{Ryosuke Sato}
\affiliation{DESY, Notkestra{\ss}e 85, D-22607 Hamburg, Germany}
%\input author_list.tex       % D0 authors (remove the first 3 lines
                             % of this file prior to submission, they
                             % contain a time stamp for the authorlist)
                             % (includes institutions and visitors)
%\date{\today}

\begin{abstract}
Sub-GeV Dark Matter particles upscattered by cosmic rays gain enough kinetic energy to pass the thresholds of large volume detectors on Earth. 
We then use public Super-Kamiokande and MiniBooNE data to derive a novel %model-independent
limit on the scattering cross section of Dark Matter with electrons
that extends down to sub-keV masses,
%\FS{[do we want to give this sub-eV info in the abstract?]},
%that extends down to $\sim 10^{-33}$~cm$^2$
closing a previously allowed wide region of parameter space. % for Dark Matter lighter than a few MeV\FS{[update with proper $\MDM$ and $\se$]}.
%, and virtually extends to arbitrarily low masses.
We finally discuss search strategies and prospects at existing and planned neutrino facilities.

\end{abstract}

\maketitle

\paragraph{\bf Introduction.}
Evidences for Dark Matter (DM) are all based on its gravitational effects, other possible interactions of this unexplained component of the Universe are currently unknown.
Some information about these interactions is obtained by \textit{direct detection} (DD) experiments, which aim at observing the scattering of DM particles off Standard Model (SM) targets~\cite{Goodman:1984dc}.
This has resulted in a huge experimental effort that, in the absence of any clear DM detection, has set strong limits on the DM-SM interactions for DM masses above few GeV, see e.g.~\cite{Akerib:2016vxi,Cui:2017nnn,Aprile:2018dbl}.

This situation is accompanied by the severe bounds that the LHC is putting on TeV-scale new physics, that cast some doubts on natural solutions to the hierarchy problem, see e.g.~\cite{Giudice:2017pzm}.
This undermines part of the motivation (i.e. the connection between naturalness and thermal relic DM) that lead to expect DM particles in the mass range where the above DD experiments are most sensitive.
It is therefore no surprise that, especially in recent years, the community has vigorously pursued the exploration of lighter DM candidates, in terms of both model building and phenomenological tests (see~\cite{Battaglieri:2017aum} for a recent report).

\medskip

The quest to determine the interactions of sub-GeV DM candidates is challenged by the low energy thresholds required by DD experiments.
Indeed, the average DM velocity $v \approx 10^{-3}$ in the Milky Way halo
 implies that sub-GeV DM induces nuclear recoils below $O$(keV), a value for which ``standard'' experiments like Xenon1T lose sensitivity. Analogously, the use of electron recoils in the same setups cannot probe DM masses below $1-10$~MeV.

A possibility to overcome this issue consists in devising new target materials and detector concepts that can be sensitive to very low-energy recoils. This direction has been widely explored in recent years, resulting in the proposal and realisation of several experiments (see again~\cite{Battaglieri:2017aum} for a review).

Another strategy to directly detect sub-GeV DM consists in relying on subdominant DM populations with much larger velocities, so that their scattering off detectors can induce energetic recoils.
A concrete example consists in ordinary DM particles upscattered in high-temperature areas of the Sun, a possibility which has been explored for DM~electron interactions in~\cite{An:2017ojc}, and for DM-nucleon ones in~\cite{Emken:2017hnp}. The internal dynamics of non-minimal dark sectors can also result in relativistic dark species, that could give signals in large detectors on Earth~\cite{Agashe:2014yua}.

\medskip

In this letter we propose a new detection strategy of sub-GeV Dark Matter, based on the subdominant component with larger kinetic energy that is unavoidably generated by cosmic rays (CRs) that scatter off DM.
Such upscattered light DM can induce visible recoils in large volume detectors, by means of the very same interactions that accelerated it.
Focusing on DM contact scatterings with electrons with cross section $\se$, we use public data of Super-Kamiokande (Super-K) and MiniBooNE to derive a new limit $\se \lesssim 10^{-(33-34)}$~cm$^2$.
This limit constitutes the strongest existing constraint on DM lighter than a few MeV, and extends to DM masses much smaller than a keV.
The possibility to probe CR interactions with light DM was first pointed out in the recent~\cite{Cappiello:2018hsu}, that derived constraints on DM from modifications of CR spectra. Our proposal tests directly the accelerated DM component by looking at its effects in detectors on Earth, rather than in CRs.

We finally discuss how searches for such a DM component could be optimised at Super-K, and the gain that one would achieve at large volume detectors with lower electron thresholds, like DUNE.
Our proposal is robust against effects that typically hamper other detection strategies of light DM, like the possible existence of other SM-DM interactions or  of small mass gaps in the dark sector.

\medskip

\paragraph{{\bf From cosmic rays to DM scatterings on Earth.}}
A diffuse flux $\phi_{i}$ of particles with a scattering cross section $\sigma_i$ with DM, of mass $\MDM$, induces a DM flux per solid angle
\beq
\frac{d \phi_\DM}{d\Omega}(K_\DM, b,l) = \frac{J(b,l)}{\MDM}\int \!\!dK_i\,\frac{d\phi_{i}}{d\Omega}(K_i) D_i^\DM
	(K_i, K_\DM)\,\sigma_i,
	\label{eq:flux_DM}
\eeq
where $J(b,l)=\int_\mathsmaller{los}\!d\ell \rho_\DM$ is the integral of the DM energy density $\rho_\DM$ over the line of sight in the direction of galactic coordinates $(b,l)$, and where we assume for simplicity that the CR flux $\phi_i$ is homogeneous inside the region of integration, which we take as customary as a cylinder centered on the galactic center (GC), with radius $R=10$~kpc and height $2h=2$~kpc. %(we will later see the impact of varying this value).
$D_i^f$ is a transfer function that encodes the energy spectrum of the particle $f$ induced by a scattering with particle $i$.
Assuming $f$ to be initially at rest in the lab frame, its final kinetic energy reads
\beq
K_f  = K_f^\text{max}\frac{1-\cos\theta}{2},\;K_f^\text{max} =  \frac{2m_f(K_i^2 + 2m_i K_i)}{(m_i+m_f)^2+2m_f K_i},
\eeq
where $\theta$ is the scattering angle in the center-of-mass (CM) frame.
If the scattering is isotropic in the CM frame, then %the distribution of $K_f$ is flat, so that
\beq
D_i^f = \frac{1}{K^\text{max}_f(K_i)}\,\Theta \left(K^\text{max}_f(K_i) - K_f \right),
\eeq
where $\Theta$ denotes the Heaviside step function. 
The number of DM scatterings with the target particles $T$ in a volume (e.g. of a detector), per time per solid angle per final energy $K_\mathsmaller{T}$ of the target particle, is then given by
\beq
\frac{dN_\DM}{dt\,d\Omega\,dK_\mathsmaller{T} } =
\int\!\!dV d K_\DM\,n_\mathsmaller{T} \sigma_\mathsmaller{T}
	\,D^{\mathsmaller{T}}_\DM(K_\DM , K_\mathsmaller{T})\,\frac{d \phi_\DM}{d\Omega},
	\label{eq:target_events}
\eeq
where $\sigma_\mathsmaller{T}$ is the scattering cross-section of DM with the target particle and $n_\mathsmaller{T}$ their number density.

\medskip
\begin{figure}[t]
\includegraphics[width=0.49\textwidth]{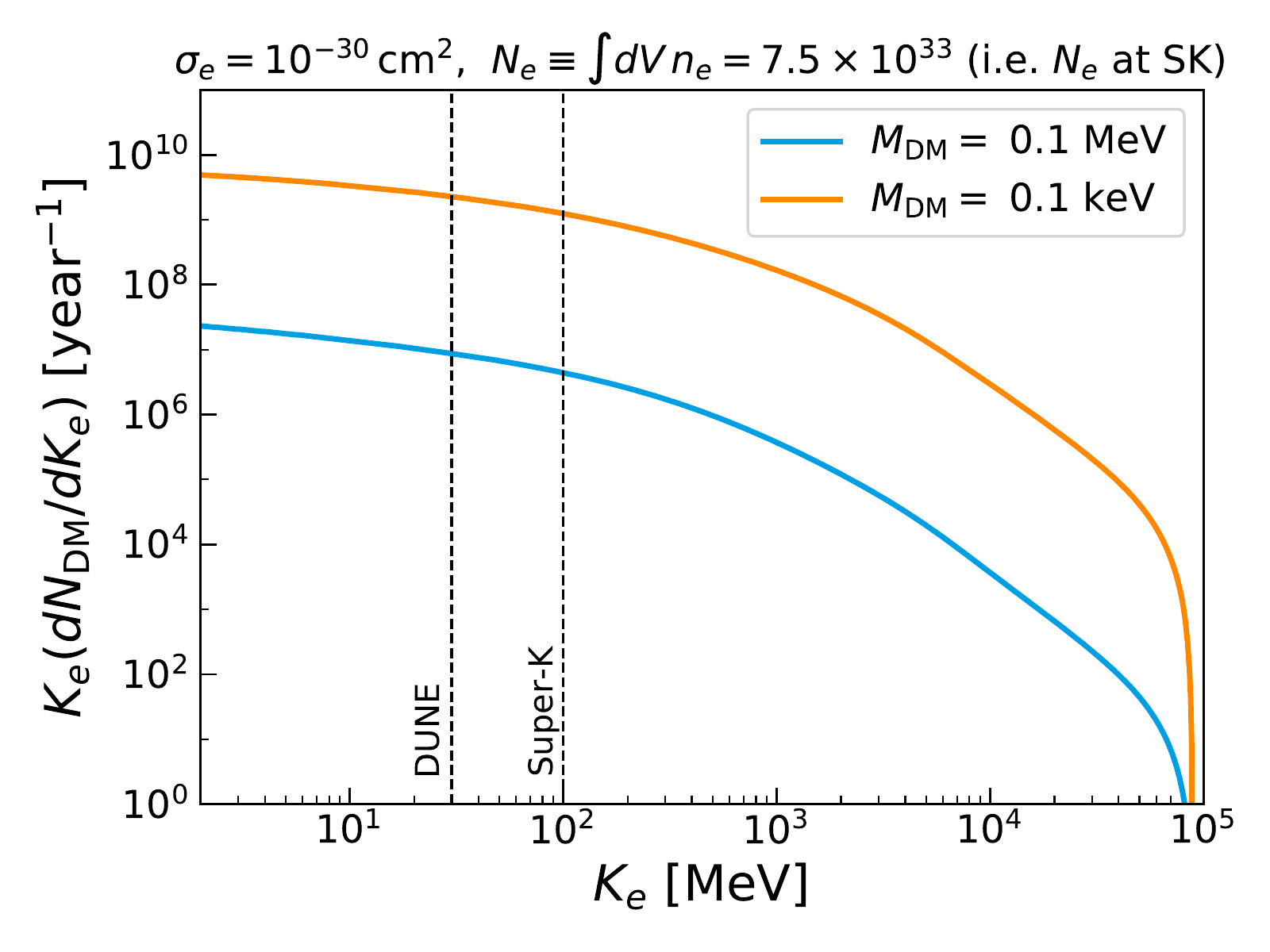}
\caption{\label{fig:electronsSK}
Kinetic energy spectrum of electrons scattered by DM. The vertical dashed lines indicate the low-energy thresholds considered at Super-K (100 MeV) and DUNE (30 MeV).
}
\end{figure}
As anticipated in the Introduction we focus on cosmic-ray electrons. We use their flux as provided in~\cite{Boschini:2018zdv} for energies between 2~MeV and 90~GeV.
To compute $J(b,l)$ we use an NFW DM density profile~\cite{Navarro:1995iw} with $\rho_\DM(r=8.5~\text{kpc})=0.42$~GeV/cm$^3$ and $r_s=20$~kpc. The precise choice of the profile has a mild impact on our treatment, because we integrate over wide areas and because the DM flux is linear in $\rho_\DM$ (analogously in a broad sense to the case of DM decay).
To give a benchmark, for $\sigma_e = 10^{-30}$cm$^2$ we find
$\phi_\mathrm{DM}(K_\mathrm{DM}=1~\mathrm{GeV}) = 4.0\times 10^{-6} (1.5\times 10^{-3})\,\mathrm{GeV}^{-1}\mathrm{sec}^{-1}\mathrm{cm}^{-2}$ for $M_\mathrm{DM} = 0.1$~MeV(keV).
See the Appendix for more details about the DM flux.
Assuming a target material containing electrons ($T=e$) and a DM-electron cross-section constant in energy results in the energy spectrum of the target electrons shown in Figure~\ref{fig:electronsSK}. %, obtained by integrating eq.~(\ref{eq:target_events}) over the whole cylinder with $R=10$~kpc and $h=1$~kpc.
Note that, once produced, DM flux propagates without any further scatterings 
as the galaxy is effectively transparent for the values of $\sigma_e$ of our interest.

An experiment that appears now in a privileged position to be sensitive to these events is Super-K, because of its unmatched large volume and because a sizable fraction of events survive the energy threshold $K_e > 100$~MeV used in current analyses (see e.g.~\cite{Kachulis:2017nci}).
As evident from Figure~\ref{fig:electronsSK}, lower $K_e$ thresholds would allow to collect more signal, but we are not aware of any existing experiment where the gain from the smaller thresholds is enough to compensate the much smaller size.
Thinking ahead, DUNE~\cite{Acciarri:2015uup} will be ideally placed to test light DM via its unavoidable relativistic component, given its expected thresholds of $K_e > 30$~MeV (see e.g.~\cite{Necib:2016aez}).

\medskip

\paragraph{\bf New constraints on light DM.}

Super-K has recently performed a search for boosted DM in its ``electron elastic scatter-like'' events with $K_e>100$~MeV~\cite{Kachulis:2017nci}, in data corresponding to 161.9 kiloton-years exposure. The results of~\cite{Kachulis:2017nci} are directly applicable to our case, as we now explain.
We use the total measured number of events reported in that paper in the first energy bin $0.1 < K_e/\text{GeV}<1.33$, $N_\text{SK} = 4042$, to place a conservative limit on light DM as
\beq
\epsilon \times N_\DM < N_\text{SK}\,,
\eeq
where $\epsilon = 0.93$ is the signal efficiency as determined in~\cite{Kachulis:2017nci}. We obtain $N_\DM$
by integrating eq.~(\ref{eq:target_events}) over the total solid angle, 2628.1 days of data-taking~\cite{Kachulis:2017nci}, and $K_e > 100$~MeV. 
We include Earth attenuation in the computation of $N_\DM$ by writing the average kinetic energy loss of a DM particle as
\beq
\frac{d K_\DM}{dz} = -n_e \se \int \!dK\,K D^e_\DM(K_\DM ,K),
\label{eq:Earth_attenuation}
\eeq
where $z$ is the depth from the Earth surface. We then assume for simplicity a constant $n_e \simeq 8\times 10^{23}$~cm$^{-3}$ (the averaged value over the Earth) and integrate 
eq.~\eqref{eq:Earth_attenuation} from $z=0$ to $z_\mathrm{SK}$ 
that is the distance between Super-K and the Earth surface 
that depends on the direction of observation ($z_\mathrm{SK} \simeq 1$~km at the zenith)), ignoring DM deflections.
We use the DM kinetic energy obtained this way in eq.~(\ref{eq:target_events}) to determine the events in the detector\footnote{An analogous treatment has been shown to be a good and conservative approximation of numerical results in~\cite{Emken:2018run} (`method b').
This is good enough for our purpose, in particular in light of the constraints we will derive from MiniBooNE.
}.

\begin{figure}[t]
\includegraphics[width=0.5\textwidth]{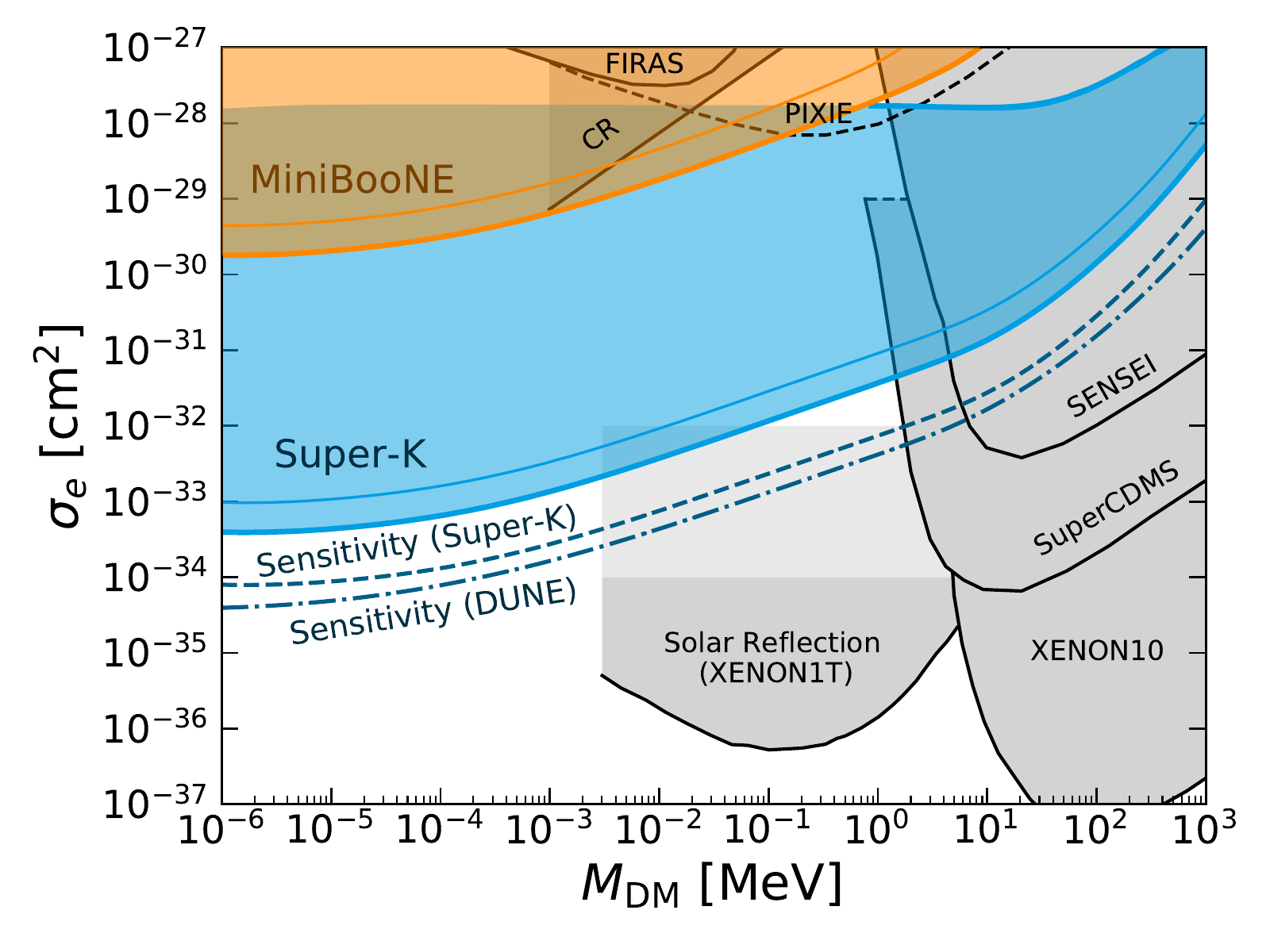}
\caption{\label{fig:limits}
Limits from Super-K (shaded blue) and MiniBooNE (shaded orange) and sensitivities at Super-K (blue dashed line) and DUNE (blue dot-dashed line) on the DM-electron scattering cross section derived in this work. They correspond to a height of the cosmic-ray electron cylinder $h=1$~kpc, the limits for a more conservative choice $h=100$~pc are shown as thin lines.
We also show
CMB anisotropies limits from FIRAS and sensitivities from PIXIE~\cite{Ali-Haimoud:2015pwa},
direct detection limits from Xenon-10~\cite{Essig:2012yx}, Super-CDMS~\cite{Agnese:2018col} and SENSEI~\cite{Crisler:2018gci}, cosmic-ray limits from~\cite{Cappiello:2018hsu}
and limits from DD of solar-reflected DM~\cite{An:2017ojc}. See text for more details.
}
\end{figure}

The resulting limit on an energy-independent $\se$ is shown as a shaded area in Figure~\ref{fig:limits}.
The even more conservative limit obtained by working with $h = 100$~pc, instead of 1~kpc, is also shown as a thin line for comparison.
The limits coming from the two higher energy bins given in~\cite{Kachulis:2017nci} result in weaker constraints than the one we show.
Our procedure sets limits in the ballpark of $\se < 10^{-33}$~cm$^2$ for $\MDM \lesssim 0.1$~keV, that slowly degrade at larger masses.

The behaviors of our exclusions can be analytically understood as follows. For 10~MeV $\gtrsim\MDM \gtrsim 0.1$~keV all cosmic rays with energy $> 100$~MeV make the Super-K electrons pass the threshold, so that the number of signal events $N_\DM$ at Super-K scales as $N_\DM \propto 1/\MDM$, following the DM number density.
Then, since $N_\DM \propto \se^2$, the excluded cross section scales $\propto \MDM^{1/2}$.
For $\MDM \lesssim 0.1$~keV the energy transferred from the CR electrons to the DM enters a regime where it is suppressed as $\MDM^{-1/2}$, because it scales as $\MDM K^2/m_e^2$.
Therefore the minimal CR energy $K_\text{min}$ required to transfer at least $\approx 100$~MeV to the DM increases at lower masses as $\MDM^{-1/2}$.
Since the CR flux scales roughly as $\phi_i \propto K^{-3}$, its integral is proportional to $K_\text{min}^{-2} \propto \MDM$. This compensates the $1/\MDM$ from the DM number density, resulting in roughly flat limits on $\se$.
For $\MDM \gtrsim 10$~MeV, the energy transferred to the electrons in Super-K scales as $m_e K^2_\DM/\MDM^2$, therefore the limit of integration in the CR energy is linear in $\MDM$. Proceeding as before we get $N_\DM \propto \se^2  \MDM^{-1} \MDM^{-2}$, where the first $\MDM$ factor is the usual consequence of the DM number density. This leads to the observed scaling of the limits as $\se \propto \MDM^{3/2}$.
%where the first $\MDM$ factor is the usual consequence of the DM number density. 
%This leads to the observed scaling of the limits as $\se \propto \MDM^{3/2}$.
As explained above, in the smallest and largest $\MDM$ regions shown in Figure~\ref{fig:limits}, the shape of our limits is driven by the CR electron of larger energies. Following~\cite{Boschini:2018zdv}, we have included their spectra only up to 90~GeV.
For more than a decade above those energies the spectral index of electrons does not become softer~\cite{Adriani:2017efm}, and this would e.g. allow to linearly extend our constraints to $\MDM$ smaller and larger than what shown in Figure~\ref{fig:limits}.

The region $\se \gtrsim 10^{-29}$ cm$^2$ that is not excluded by Super-K is accessible at surface neutrino detectors\footnote{See~\cite{Kim:2018veo} for a recent list of such experiments with references, and~\cite{Chatterjee:2018mej} for a study of boosted DM at proto-DUNE.
}.
To demonstrate this point, we use the MiniBooNE measurement~\cite{Aguilar-Arevalo:2018wea} of 2 events of $\nu-e$ scattering, in a region defined by $\cos\theta_e > 0.9$ along the line between the detector and the neutrino beam, and by $75 < K_e/\text{MeV} < 850$.
DM accelerated by CR electrons induce a number of electron scatterings at MiniBooNE that we 
%determine using
compute using eq.~(\ref{eq:target_events}) with the same energy and angular cuts, and a volume of 139 tons ($N_e \simeq 5\times 10^{31}$) that we infer from~\cite{Aguilar-Arevalo:2018wea} as the one contained in a radius of 3.38~meters (we conservatively interpret the $\nu-e$ cut `distance to wall' as referring to the distance from the spherical optical separation).
We then integrate over a time of 124 seconds, that we obtain multiplying the observation time per pulse of 2~$\mu$s (third cut in Table~III of~\cite{Aguilar-Arevalo:2018wea}), with the total number of triggered pulses $6.2 \times 10^7$.
The latter is not explicitly given in~\cite{Aguilar-Arevalo:2018wea}, but we infer it as the total number of protons on target ($1.86\times 10^{20}$) divided an average number of protons per pulse of $3\times 10^{12}$ that again we infer from~\cite{Aguilar-Arevalo:2018wea}.
We include the Earth attenuation using eq.~(\ref{eq:Earth_attenuation}), where the amount of crust that DM goes across depends on $\theta_e$, the azimuthal angle $\phi_e$ and the depth of the booster $\simeq 6$~meters~\cite{FNAL:booster}.
For simplicity we conservatively take the same value for the depth of MiniBooNE, corresponding to that of its center~\cite{AguilarArevalo:2008qa}.
We finally multiply the signal events by 0.15 (signal efficiency inferred from~\cite{Aguilar-Arevalo:2018wea}) and impose the result to be smaller than the observed 2 events.
The resulting constraint is displayed in Figure~\ref{fig:limits}. It extends to $\se \gtrsim 10^{-27}$~cm$^2$, that we do not show as that would require a treatment of DM scattering through the atmosphere, which goes beyond the purpose of this paper.
The analysis of more MiniBooNE data should allow to close the small gap between the Super-K and MiniBooNE exclusions
at $\MDM \gtrsim 1$~MeV.
Our conservative MiniBooNE analysis, while admittedly rough, clearly demonstrates the point that cross sections larger than $10^{-29}$~cm$^2$ are accessible at surface neutrino detectors.

\medskip

\paragraph{\bf Sensitivities at Super-K and DUNE.}

We estimate them using the signal spatial information, i.e. the larger number of signal events expected from the direction of the galactic center.
We integrate the signal over a cone with axis centered on the direction of the GC and opening angle of $10^\circ$, corresponding to the opening angle from Earth of the height of the cylinder assumed to contain the CR electrons, $h=1$~kpc.
In an actual search at neutrino experiments, the background could be estimated at Super-K using part of the space complementary to the cone as a control-region, similarly to what has been done in~\cite{Kachulis:2017nci}.
The uncertainty on the background would then be dominated by statistics, so that we  estimate a 95\%CL reach on light DM by imposing
\beq
\left.\frac{N_\DM}{\sqrt{N_\DM+N_\text{bkg}}}\right|^{10^\circ}_\text{a.h.} = 2\,. 
\eeq
The subscript refers to the fact that we only use the fraction of the events above horison, to be conservative with respect to the attenuation of the DM flux from Earth crossing.

In practice, we determine $N_\text{bkg}^\text{SK}$ at Super-K by multiplying the total events measured in the first energy bin~\cite{Kachulis:2017nci} by the fraction of the sky over which we integrate $\simeq 0.01$, i.e. using the observed isotropy of the background. We determine $N_\text{bkg}^D$ at DUNE assuming 200 kton-year of data (to have the same number of electron-year of Super-K), and using $\left.dN_\text{bkg}^D/dt\right|_{10^\circ} = 0.1$~event/kton-year~\cite{Necib:2016aez}. We finally multiply the Super-K (DUNE) background events by 0.37 (0.32), i.e. by the time the GC is above the horizon, that we determine with~\cite{astropy:2013,astropy:2018}. For the signal, we integrate eq.~(\ref{eq:target_events}) over the above cone (the signal fraction surviving is $\simeq 0.15$) and multiply by 0.37 (0.32) at Super-K (DUNE).
Other large-volume detectors, like Hyper-K, have also promising sensitivities that can be determined as above.

The results are displayed in Figure~\ref{fig:limits}. The smallest values of the cross sections to which both Super-K and DUNE are sensitive to are such that the Earth would be actually transparent to DM. This would allow, when performing an actual search, to gain sensitivity both from using events under the horison, and by performing a full optimization of the region of integration (which we expect would have a wider opening angle in the direction of the galactic plane).

\medskip

\paragraph{\bf Other light-DM searches.}

In Figure~\ref{fig:limits} we also display constraints
from a variety of searches:

\begin{itemize}
\item[$\diamond$] direct detection constraints from Xenon-10~\cite{Essig:2012yx} and Super-CDMS~\cite{Agnese:2018col}, that we stop at $\se = 10^{-29}$~cm$^2$ to conservatively account for the $O(1)$~km of rock above the detectors, and because larger cross sections are anyway probed by the SENSEI surface run~\cite{Crisler:2018gci};
\item[$\diamond$] constraints from CMB anisotropies from the FIRAS experiment, and related projections at PIXIE~\cite{Ali-Haimoud:2015pwa};
\item[$\diamond$] constraints from the observed cosmic-ray electron spectra~\cite{Cappiello:2018hsu}, that to be conservative we do not extend below the $\MDM$ range given in~\cite{Cappiello:2018hsu}, because there the kinematical regime driving the shape of the line changes;
\item[$\diamond$]  The Xenon1T constraints induced by the population of DM reflected by the core of the Sun~\cite{An:2017ojc}, whose large temperature can provide the DM with enough kinetic energy to pass the thresholds of DD experiments on Earth.
\end{itemize}

The Sun constraints
are given in~\cite{An:2017ojc} up to $\se = 10^{-34}$~cm$^2$, and down to $\MDM = 3$~keV.
We do not show them for $\MDM < 3$~keV because, in that range,
the simple one-scattering regime with the core of the Sun is not enough to give the target electrons in the detectors enough energy to pass the cut of $0.19$~keV used in~\cite{An:2017ojc}. Therefore the study of those masses requires a treatment that goes beyond the purpose of this letter.
We also do not extend these limits above $\se = 10^{-32}$~cm$^2$, because they make the radial extension of the radiative area of the Sun become much larger than the related DM-electron interaction lengths, $R_\text{rad} \simeq 0.5 R_\text{sun} \gg (\se n_e)^{-1}$, where e.g. $n_e \approx 10^{23}$~cm$^{-3}$ at the edge between the radiative and convective areas~\cite{NASA:sun}.
Therefore DM particles are expected to scatter several times in the radiative and convective regions, whose temperatures are much smaller than in the core of the Sun, leading to the expectation that the limits of~\cite{An:2017ojc} will be strongly affected.\footnote{Fig.~3
of~\cite{An:2017ojc} indeed shows that for $\se = 10^{-33}$~cm$^2$ the maximal DM energy is smaller than for smaller cross sections.}. A more precise determination of this effect goes beyond the purposes of this paper.
This obstruction might be less severe for the SENSEI~\cite{Crisler:2018gci} and Super-CDMS~\cite{Agnese:2018col} sensitivities to DM reflected from the Sun, shown in~\cite{An:2017ojc}. However, in the absence of a detailed simulation of propagation of DM in the Sun and of its effects on such detectors, we refrain from showing those sensitivities in our plots.

We finally remark that, in presence of additional interactions with the SM (e.g. with nucleons), the physics of DM escaping the Sun will become even more dependent on the outer Sun layers. Our limits from Super-K are instead more robust against assuming such an extra interaction (they would actually improve thanks to the extra upscattered component from cosmic-ray protons), until it prevents DM from reaching the detector.

We do not show limits on $\se$ coming from the combination of CMB and BBN data~\cite{Boehm:2013jpa,Nollett:2013pwa,Nollett:2014lwa}, as they may be attenuated or evaded depending on other model assumptions, like the existence of additional dark radiation or annihilation channels for DM.
Analogously, we do not show CMB constraints on annihilating DM, as they are more model-dependent and for example they are weak if DM annihilation is $p$-wave (see e.g.~\cite{Liu:2016cnk}).

\medskip

\paragraph{\bf On concrete light-DM models.}
A plethora of models of sub-GeV DM and dark sectors have recently been proposed: just to name a few SIMPs~\cite{Hochberg:2014dra,Choi:2017zww}, ELDERs~\cite{Kuflik:2015isi}, light dark sectors and/or DM from supersymmetry~\cite{ArkaniHamed:2008qp}, from leptogenesis~\cite{Falkowski:2017uya}, from the hierarchy problem~\cite{Fonseca:2018kqf,Banerjee:2018xmn}, or demanded by observed anomalies, e.g. in $B$ decays~\cite{Sala:2017ihs,Sala:2018ukk}.
Inspired by this rich model-building activity, we now briefly comment about the application of our results to some concrete models of light DM. A more detailed exploration of the following and other applications, while certainly interesting, goes beyond the purpose of this letter.

An explicit example for which our strategy looks particularly promising is that of dark sectors with small mass splittings, see e.g.~\cite{Kim:2016zjx,Darme:2017glc,Giudice:2017zke,Darme:2018jmx}. These models can have sizeable DM-electron interactions while evading limits from cosmology, SENSEI, Super-CDMS etc. because in these energy domains the DM-electron scattering is inelastic. Our proposal avoids that limitation thanks to its larger energy regimes, and therefore stands out as a prominent possibility to directly test such DM candidates.

We also studied for simplicity energy-independent contact interactions. The impact of these searches to other regimes can be grasped by observing that the energy exchanges that drive our sensitivities are of the order of the threshold of the neutrino detectors, $K_e >30$--$100$~MeV.
Therefore the performance of our proposal, with respect to other DD probes that rely on smaller energy exchanges (Sun reflection, CMB, Super-CDMS etc.), would be better than what displayed in Figure~\ref{fig:limits} if $\sigma_e$ grows with increasing energy (e.g. as in the case of SM neutrinos), and would be worse in the opposite case (e.g. for mediators much lighter than $O(100)$~MeV, see e.g.~\cite{Barkana:2018qrx}).

Finally, if the relic particle $\chi$ interacting with electrons constitutes a subdominant component of DM, $f=\Omega_\chi/\Omega_\mathsmaller{DM} <1$, then our constraints and sensitivities on $\se$ are relaxed by $\sqrt{f}$, unlike the more severe rescaling by $f$ of other DD probes.

\medskip

\paragraph{{\bf Conclusions and Outlook.}}
The results presented in this letter demonstrate that large-volume neutrino experiments have a promising potential to probe unexplored regimes of light-DM interactions with the SM. This physics case relies on our novel proposal to test the energetic DM component that is unavoidably generated by scatterings with CR electrons in the galaxy.
The conservative limit we set using public Super-K data excludes previously allowed wide regions of parameter space, and that could be improved if a dedicated search would be performed in existing data at Super-K, see Figure~\ref{fig:limits}. The prospects of other large neutrino experiments, like Hyper-K and DUNE, also look bright.

Thinking about possible future directions, going to lower electron energy thresholds would increase the signal by allowing to be sensitive to a  larger fraction of the upscattered DM (see Figure~\ref{fig:electronsSK}). 
That would pose the challenge of dealing with much larger backgrounds, e.g. from solar neutrinos~\cite{Gutlein:2010tq}. While we do not explore this regime further here, we encourage the experimental collaborations to pursue that direction, for example by employing the peculiar modulation of the signal (from the daily rotation of the GC direction).

\medskip

\subsection{Acknowledgements}

%\acknowledgements
We thank Kfir Blum and Luc Darm\'e for useful discussions and Christopher Cappiello for pointing out a numerical glitch in our previous version of Figures~1 and~2, which does not affect our conclusions.
\medskip

{\footnotesize
\noindent Funding and research infrastructure acknowledgements: 
\begin{itemize}
\item[$\ast$] Y.E. is supported in part by a JSPS KAKENHI Grant No. JP18J00540;
\item[$\ast$] F.S. is supported in part by a {\sc Pier} Seed Project funding (Project ID PIF-2017-72).

\end{itemize}
}

\subsection{Note added}
{\footnotesize
When this work was in preparation, ref.~\cite{Bringmann:2018cvk} appeared proposing the same idea that DM upscattered by cosmic rays can give observable effects in Earth detectors.  That work is complementary to ours in that it focuses on DM-nucleon interactions and on signals at detectors like Xenon-1T, while we focus on DM-electron interactions and on signals at large neutrino experiments. }

%%%%%%%%%%
\appendix
%%%%%%%%%%
\section{Appendix}

We discuss here the kinetic energy distribution of the dark matter
and the effects of the Earth attenuation.

\medskip

\paragraph{\bf Dark matter kinetic energy distribution.}

\begin{figure}[t]
\centering
\includegraphics[width=0.9\hsize]{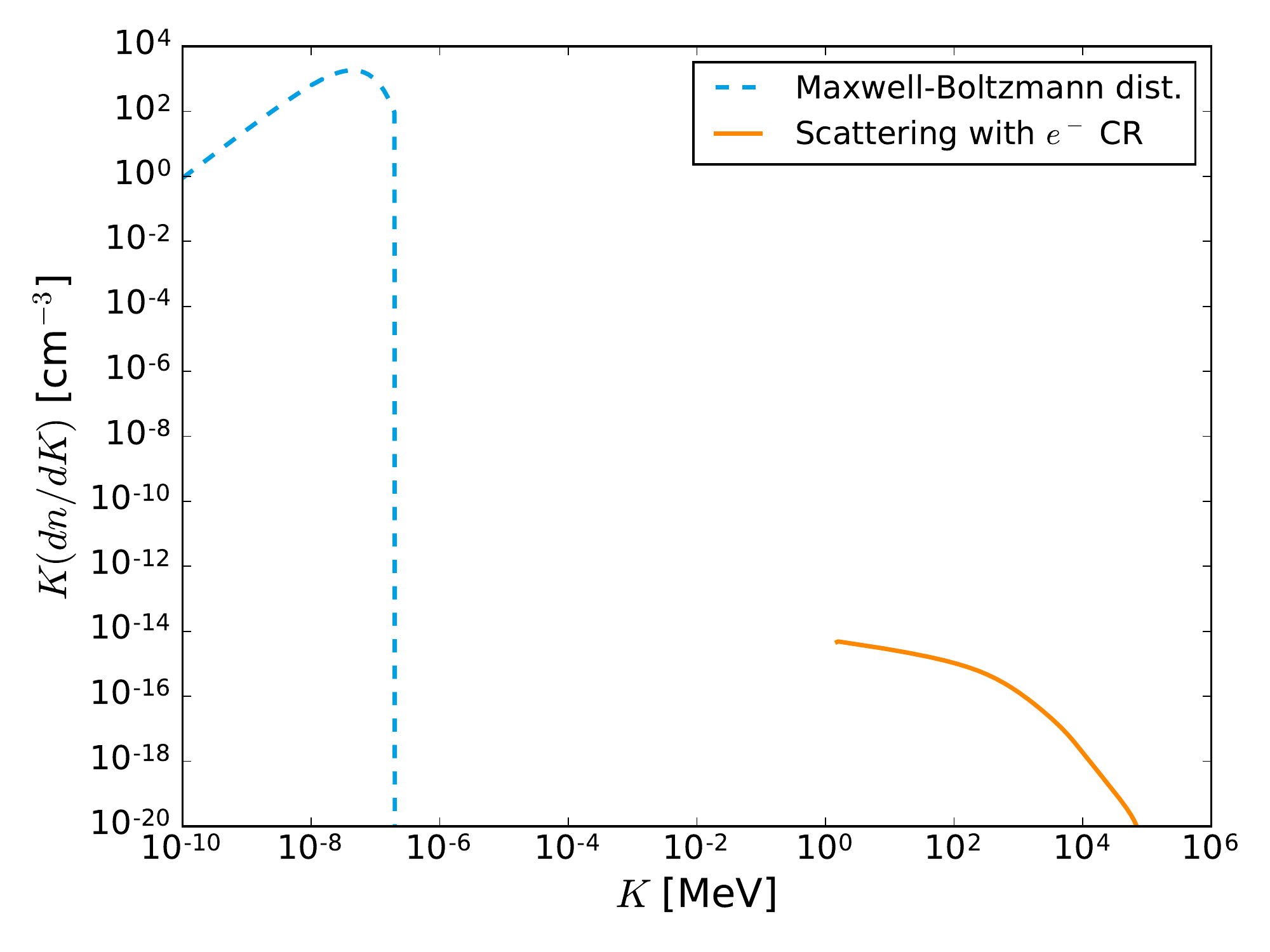}
\caption{The kinetic energy distribution of the dark matter with $M_{\rm DM} = $ 0.1\,MeV.
The blue line shows the standard halo dark matter distribution, and the orange line shows the one of the dark matter boosted by scattering of cosmic ray electrons, for $\sigma_e = 10^{-30}$\,cm$^2$.
}\label{fig:100keV}
\end{figure}

\begin{figure}[h]
\vspace{5mm}
\includegraphics[width=0.9\hsize]{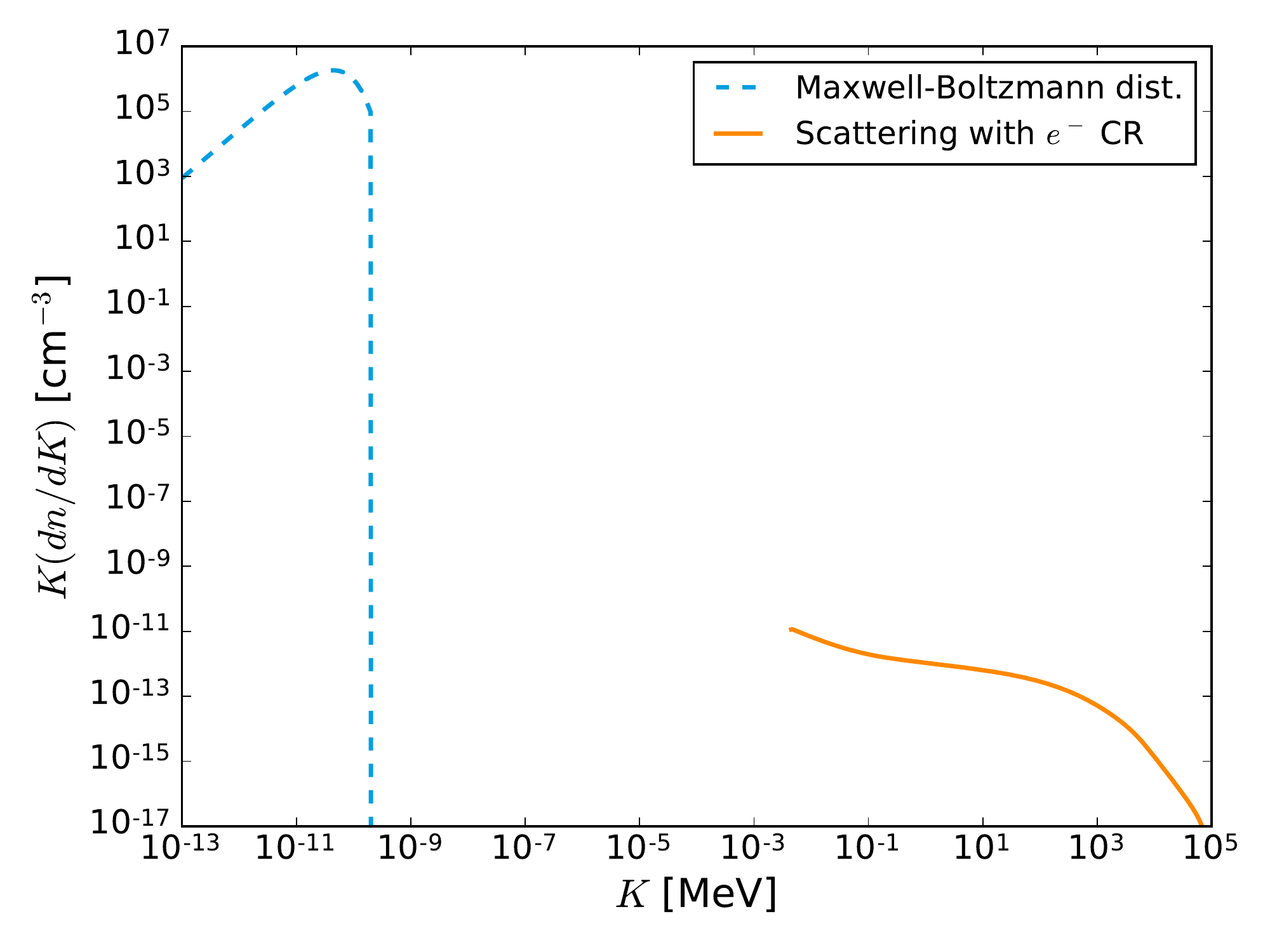}
\caption{Same as Fig.~\ref{fig:100keV} with $M_{\rm DM} = $ 0.1\,keV.}\label{fig:100eV}
\end{figure}

Here we compare the kinetic energy distribution of the dark matter component which is boosted by cosmic ray electrons with that of the standard halo dark matter.
Although the non-relativistic component of the halo dark matter is roughly described by a Maxwell-Boltzmann distribution,
the behaviour of its high energy tail has been confirmed by neither observations nor $N$-body simulations.
%%%%%%%%%%%%%%%%%%%%%%%%%%%%%%%%%%%%%%%%%%%%%
%In this note, we simply consider a Maxwell-Boltzmann distribution with mean velocity of 100\,km/s \cite{Kuhlen:2009vh}.
%%%%%%%%%%%%%%%%%%%%%%%%%%%%%%%%%%%%%%%%%%%%%
In this note, we simply consider a truncated Maxwell-Boltzmann distribution \cite{Lewin:1995rx, Lisanti:2016jxe}, which is given as
\begin{align}
\frac{dn}{dv} &= \frac{4}{N\sqrt{\pi}} \frac{v^2}{v_0^3} \exp\left( -\frac{v^2}{v_0^2} \right) \theta(v_{\rm esc} - v), \\
N &= {\rm erf}\left( \frac{v_{\rm esc}}{v_0} \right) - \frac{2}{\sqrt\pi} \frac{v_{\rm esc}}{v_0} \exp\left( -\frac{v_{\rm esc}^2}{v_0^2}\right).
\end{align}
Here we take the most probable speed $v_0$ as 230\,km/s and the escape velocity $v_{\rm esc}$ as 600\,km/s.
The normalization factor $N$ is determined to satisfy $\int dv (dn/dv) = 1$.
(For deviation from Maxwell-Boltzmann distribution, see \textit{e.g.,} Ref.~\cite{Kuhlen:2009vh}.)
%%%%%%%%%%%%%%%%%%%%%%%%%%%%%%%%%%%%%%%%%%%%%
%We also assume the local dark matter density is $0.4$\,GeV/cm$^{-3}$.
For the boosted component, we assume a DM-electron scattering cross section $\sigma_e = 10^{-30}$\,cm$^2$, and a height of the CR cylinder $2h=2$\,kpc.
We cut the lower energy component because, to determine it, we should know the low energy $e^-$ flux, which is not provided below 2 MeV in~\cite{Boschini:2018zdv}.
We make the comparison for $M_{\rm DM} = 0.1$\,MeV in Fig.~\ref{fig:100keV},
and $M_{\rm DM} = 0.1$\,keV in Fig.~\ref{fig:100eV}.

\medskip

\begin{figure}[t]
\centering
\includegraphics[width=0.9\hsize]{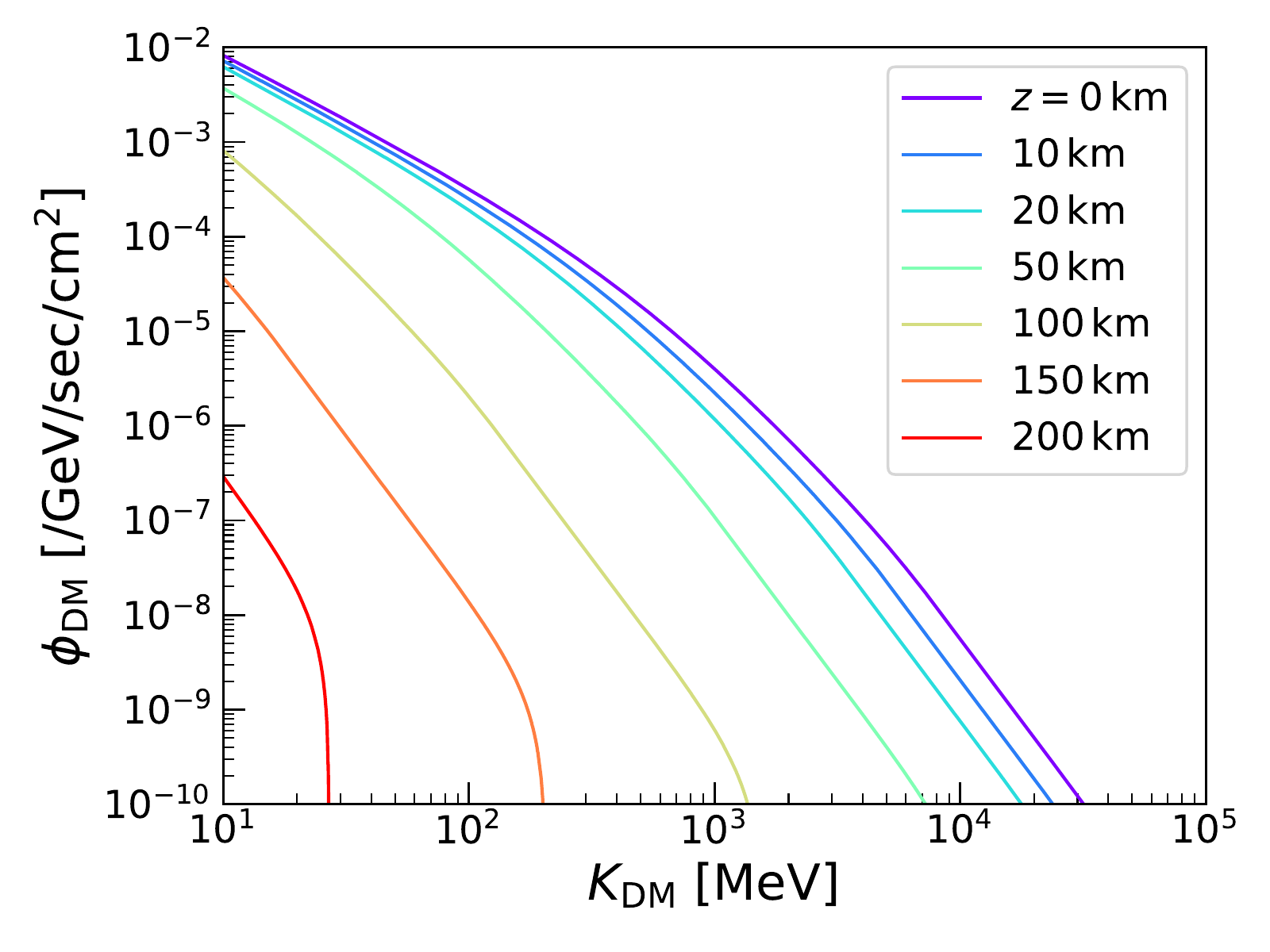}
\caption{The effects of the Earth attenuation on the dark matter flux for $M_{\rm DM} = $ 0.1\,MeV.
The depth $z$ measures the distance from the Earth's surface.
The other parameters are taken as $\sigma_e = 10^{-30}\,\mathrm{cm}^2$, $R = 10\,\mathrm{kpc}$
and $h = 1\,\mathrm{kpc}$.}
\label{fig:flux_atten1}
\end{figure}

\begin{figure}[h]
\centering
\vspace{5mm}
\includegraphics[width=0.9\hsize]{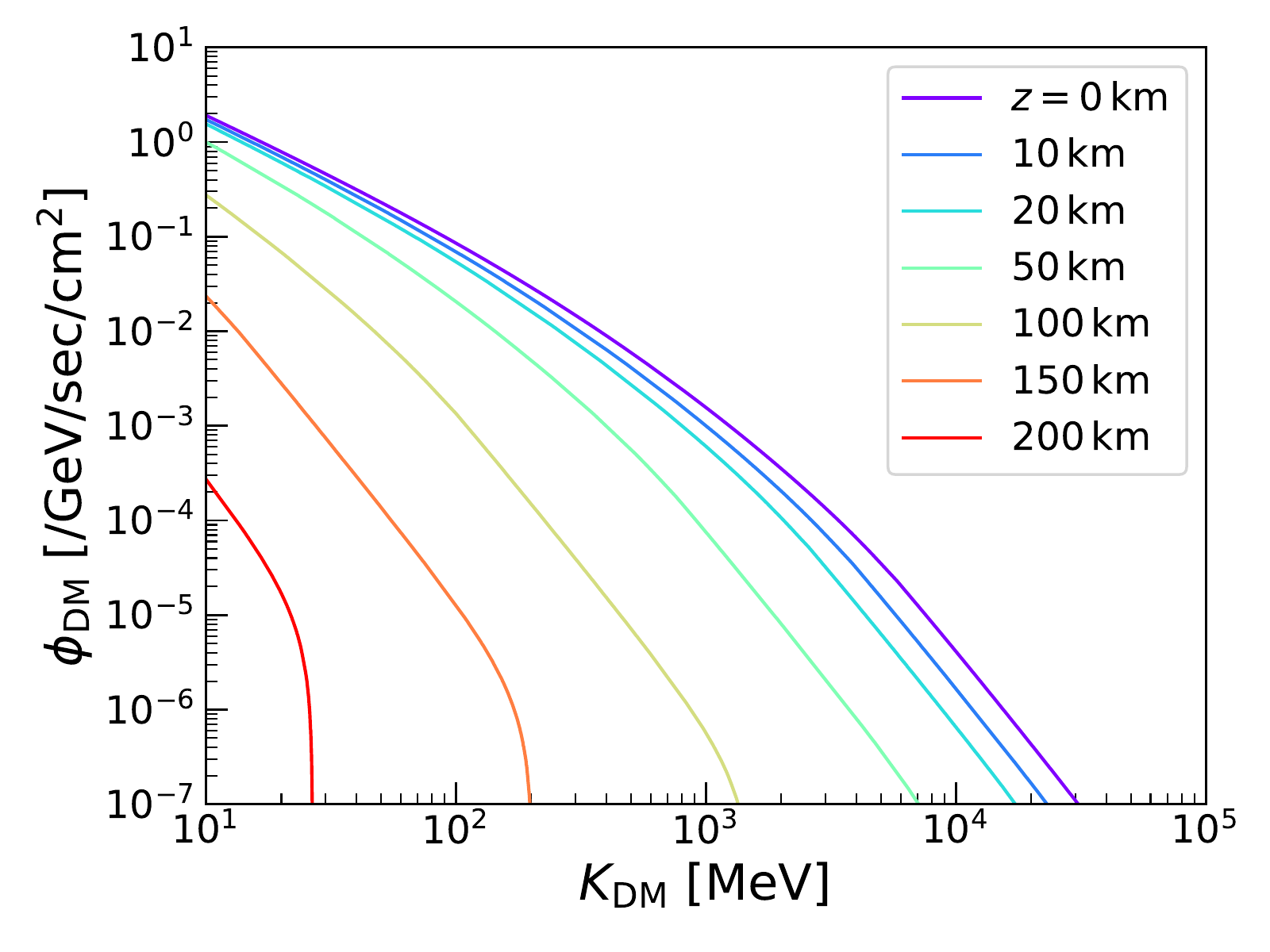}
\caption{Same as Fig.~\ref{fig:flux_atten1} with $M_{\rm DM} = $ 0.1\,keV.}\label{fig:flux_atten2}
\end{figure}

\paragraph{\bf Effects of the Earth attenuation.}
In Figs.~\ref{fig:flux_atten1} and~\ref{fig:flux_atten2},
we show the dark matter flux for a different value of the depth $z$
for $M_\mathrm{DM} = 0.1\,\mathrm{MeV}$ and $0.1\,\mathrm{keV}$,
$\sigma_e = 10^{-30}\,\mathrm{cm}^2$,
$R = 10\,\mathrm{kpc}$ and $h = 1\,\mathrm{kpc}$.
It is computed from the DM flux on the surface by the relation
\begin{align}
	dK_\mathrm{DM}(z) \phi_\mathrm{DM}(K_\mathrm{DM}(z),z)
	= dK_\mathrm{DM} \phi_\mathrm{DM}(K_\mathrm{DM}),
	\label{eq:number_conservation}
\end{align}
where the quantities without $z$-dependence are defined on the Earth's surface.
Eq.~\eqref{eq:number_conservation} follows from the number density conservation of the DM within
our approximation.
It is clear from the figures that the DM flux is attenuated a lot once $n_e \sigma_e z \gtrsim 1$.
Actually the Earth attenuation depends on $\sigma_e$ and $z$ only through the combination $n_e \sigma_e z$,
and hence one can deduce a similar behaviour for different values of $\sigma_e$ and $z$.


\bibliography{LightDM_Neutrinos}

\end{document}